\documentstyle[aps,prl,epsf,floats]{revtex}

\def\etal{{\it et al}}
\def\X{$\times$}
\def\a{{\bf a}}
\def\b{{\bf b}}
\def\c{{\bf c}}
\def\d{{\bf d}}
\def\e{{\bf e}}
\def\sop{\mbox{$S_1^+$}}
\def\som{\mbox{$S_1^-$}}
\def\stp{\mbox{$S_2^+$}}
\def\stm{\mbox{$S_2^-$}}

\begin{document}

\wideabs{ \title{Theory of the ``Honeycomb Chain-Channel''
Reconstruction of Si(111)3\X 1}

\author{Steven C. Erwin}
\address{Complex Systems Theory Branch, Naval Research Laboratory,
Washington, D.C. 20375}

\author{Hanno H. Weitering}
\address{Department of Physics and Astronomy, The University of
Tennessee, Knoxville, Tennessee 37996\\and Solid State Division, Oak
Ridge National Laboratory, Oak Ridge, Tennessee 37831}

\date{\today}
\maketitle
\begin{abstract}
First-principles electronic-structure methods are used to study a
structural model for Ag/Si(111)3\X 1 recently proposed on the basis of
transmission electron diffraction data. The fully relaxed geometry for
this model is far more energetically favorable than any previously
proposed, partly due to the unusual formation of a Si double bond in
the surface layer. The calculated electronic properties of this model
are in complete agreement with data from angle-resolved photoemission
and scanning tunneling microscopy.
\end{abstract}




}

The surfaces of silicon reconstruct in strikingly diverse ways. This
diversity provides a rich proving ground for simple, physically
intuitive ideas about the stability of semiconductor surfaces---ideas
which are invaluable for understanding more complex dynamical
phenomena such as growth, etching, and reactivity. Two such simple
concepts---elimination of surface dangling bonds and relief of surface
stress---explain the frequent appearance of elementary ``building
blocks'' in silicon reconstructions. For example, dimers appear on
Si(001)2\X 1 (and its variants), (111)7\X 7, (113)3\X 2, (114)2\X 1,
and (5,5,12)2\X 1; $\pi$-bonded chains appear on (111)2\X 1 and
(5,5,12)2\X 1; and adatoms appear on the (111)7\X 7, 
(113)3\X 2, (114)2\X 1, and (5,5,12)2\X 1 reconstructions (see
Ref.\ \cite{baski97} and citations therein).

One reconstruction that remains controversial is the metal-induced
$M$/Si(111)3\X 1 (where $M$=Li, Na, K, Ag, Mg), which is widely believed
to have a single common structure.  Starting from known building
blocks, we and others have recently proposed two models for this
reconstruction, both with low dangling-bond density and low surface
stress \cite{okuda94,sakamoto94,wong94,weitering94,erwin95}.
First-principles total-energy calculations showed these models to be
stable relative to previously proposed ones \cite{erwin95}. The more
stable of the two, the extended Pandey chain model, was also
consistent with scanning tunneling microscopy (STM) images, but the
calculated surface-state band structure of both models was in serious
disagreement with angle-resolved photoemission (ARPES) data
\cite{weitering96}.

In this Letter, we examine theoretically a new model for the $M$:3\X 1
surface proposed very recently by Collazo-Davila, Grozea, and Marks
(CGM) on the basis of direct phasing of transmission diffraction data
\cite{collazo98}, and independently by Lottermoser \etal.\ from
surface x-ray diffraction and total-energy calculations
\cite{lottermoser98}.  First, starting from coordinates obtained by
CGM, we further relax the atomic positions so as to minimize the
calculated total energy. The resulting model is by far the most stable
of any proposed to date. Second, we show how a nearly perfect
``surface symmetry'' of this reconstruction neatly resolves a
long-standing puzzle regarding ARPES data for Li:3\X 1 (and may also
explain the apparent insulating nature of Mg:3\X 1, which has an odd
number of electrons per unit cell). Third, we show that this model
derives its remarkable stability from the formation of a true Si
double bond---a ``building block'' not seen on any other surface of
silicon. Finally, we show that this model completely accounts for the
appearance of the existing STM images (including the
differences between $M$=Li and Ag).

The model proposed by CGM was based on data from Ag:3\X 1
\cite{collazo98}. Previous experimental results from low-energy
electron diffraction \cite{fan90}, STM \cite{wan92}, and core level
spectroscopy \cite{okuda94,weitering96} suggest that Li, Na, K, Ag,
and Mg all have the same reconstruction (apart from several details
discussed later). For this reason we concentrate first on Li:3\X 1,
which leads to a slightly simpler physical picture, and then later
discuss the differences between Li:3\X 1 and Ag:3\X 1.  Starting from
the atomic positions determined by CGM, we performed full structural
relaxation using first-principles electronic-structure methods.  The
calculations used a double-sided slab geometry with 14 layers of Si
plus Li, with a vacuum region equivalent to 10 layers of Si.  The four
innermost layers of Si were fixed, while all other atoms were allowed
to relax until the rms force was less than 0.02 eV/\AA.  Total
energies and forces were calculated within the local-density
approximation (LDA), using Troullier-Martins pseudopotentials and a
plane-wave basis with a kinetic-energy cutoff of 12 Ry, as implemented
in the {\sc fhi96md} code \cite{bockstedte97}.  Four $k$-points were
used to sample the surface Brillouin zone.

\begin{figure}[tbp]
\epsfxsize=5.5cm\centerline{\epsffile{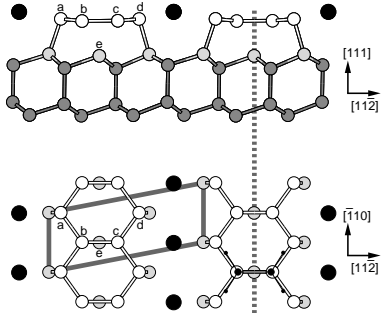}}
\caption{Fully relaxed geometry for the honeycomb chain-channel 
model of Li/Si(111)3\X 1. Black circles are Li. White, light gray, and
dark gray circles are the surface layer, first layer, and deeper Si
layers, respectively. The heavy dotted line is an approximate
mirror-symmetry plane for the top two Si layers (see discussion in
text). The 3\X 1 surface unit cell is shown as a heavy solid line.  A
disilene molecule, Si$_2$H$_4$, is shown in the lower right (see
discussion).
\label{model}}
\end{figure}

The fully optimized structure is shown in Fig.~\ref{model}. Several
observations will motivate our discussion below. (1) There are four
inequivalent Si surface atoms (labeled \a, \b, \c, \d\ in
Fig.~\ref{model}) which form a ``honeycomb chain'' lying in a plane
parallel to the surface. (2) Each of these surface Si atoms is
threefold coordinated: outer atoms \a\ and \d\ roughly tetrahedrally,
and inner atoms \b\ and \c\ in a planar configuration.  (3) Electron
density plots show that \b\ and \c\ are only very weakly bonded 
to the first-layer Si atom, \e,
below. (4) The Li atom sits in a channel formed by neighboring
honeycomb chains, and occupies a symmetric location with respect to
the [$\overline{1}$10] direction---that is, it preserves the
[$\overline{1}$10] mirror plane.  (5) Electron density plots show that
the Li atom is fully ionized.

To compare the stability of this model with previous ones we compute
the surface energy, $E_s=[E_t(N)-NE_t^{bulk}]/2$, where $E_t(N)$ is
the total energy of a reconstructed slab supercell containing $N$ Si
atoms, and $E_t^{bulk}$ is the total energy per atom of bulk Si.
Relative to the extended Pandey model, which is the lowest energy
model to date \cite{erwin95}, this ``honeycomb chain-channel'' (HCC)
model has a surface energy lower by 0.52 eV per 3\X 1 cell, or 14
meV/\AA$^2$---an enormous energy gain on the scale of silicon surface
reconstructions (and in good agreement with the LDA result of Ref.\
\cite{lottermoser98}).

A structural model with five under-coordinated atoms
would appear to be energetically unfavorable, so how can we
understand the very low calculated surface energy of the HCC model?
We develop here a simple intuitive picture for this stability, one
which also resolves an apparent discrepancy regarding ARPES data for
Li:3\X 1. The measured spectra reveal a {\it single surface state},
dispersing downward from $\Gamma$ toward the A point, i.e., along the
[$\overline{1}$10] direction \cite{weitering96}. The apparent absence
of additional states is puzzling since, roughly speaking, each surface
orbital is expected to give rise to a surface state---and the 3\X 1
structural models proposed to date have at least three surface
orbitals.  Indeed, the HCC model has five unsaturated Si surface
orbitals, and so should give rise to five surface states. There are
six electrons available to fill these states---four from the Si surface
atoms (\a, \b, \c, \d), one from the Li ion, and one from \e---and
so we expect three filled surface states and two empty states.

\begin{figure}[tbp]
\epsfxsize=6.5cm\centerline{\epsffile{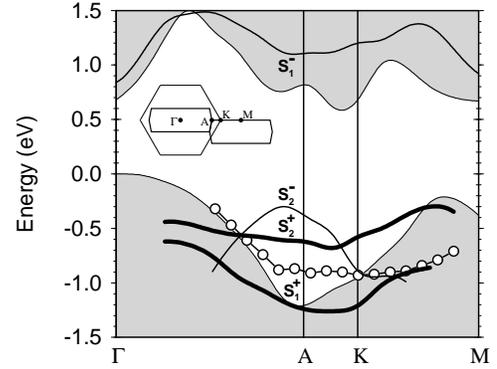}}
\caption{Theoretical band structure for the honeycomb chain-channel model
of Li:3\X 1. Surface states, $S$, are denoted by solid curves: heavy
lines for states of even parity, and light lines for states of odd
parity. The projected 3\X 1 bulk bands are shaded. Experimental
angle-resolved photoemission peak positions are indicated by circles.
\label{surfacebands}}
\end{figure}

The theoretical band structure, part of which is shown in
Fig.~\ref{surfacebands}, confirms this description and illustrates the
discrepancy. Three surface states, labeled $S_1^+, S_2^+, S_2^-$, are
fully occupied and one, labeled \som, is completely empty (the other
empty state lies higher and is not shown).  For the occupied states,
the single peak in the ARPES data, also shown in
Fig.~\ref{surfacebands}, lies between and thus is consistent with
\sop\ and \stp, but is completely inconsistent with \stm.

The resolution to this discrepancy provides internally consistent
evidence in support of the HCC model itself, and also explains its
surprising stability.  Our argument is that the photoemission
cross-section is large for states \sop\ and \stp, but small or
vanishing for \stm. Specifically, if the plane defined by the incident
light beam and the photoelectron detector is a mirror plane
of the crystal, then the states along that direction in
reciprocal space have either even or odd parity. Given this condition,
the photoemission matrix element, $\langle \Psi_i | {\bf A}\cdot {\bf
p}| \Psi_f \rangle$, will vanish by symmetry for two situations: (i)
when the polarization vector, {\bf A}, is normal to the mirror plane
and the initial wave function, $\Psi_i$, has even parity; or (ii) when
{\bf A} is parallel to the mirror plane and $\Psi_i$ has odd parity
\cite{plummer82}.

The ARPES data shown in Fig.~\ref{surfacebands} are from a synchrotron
light source with the incident beam, detector, and polarization
vector all lying in the plane defined by the surface normal and the
[$\overline{1}$10] direction, i.e. the plane normal to
[11$\overline{2}$]. This plane is shown in Fig.~\ref{model} as a heavy
dotted line. It is clearly {\it not} a mirror plane with respect to
either the Li atoms or the Si substrate. However, the Li atoms are
completely ionized, and the substrate atoms do not contribute
significantly to the surface states. Hence neither is relevant to
determining the surface symmetry.  The relevant atoms are those in the
surface layer (\a, \b, \c, \d) and perhaps the first layer.  
Fig.~\ref{model} shows that for these atoms the dotted
line is a nearly perfect mirror plane.

\begin{figure}[tbp]
\epsfxsize=7.5cm\centerline{\epsffile{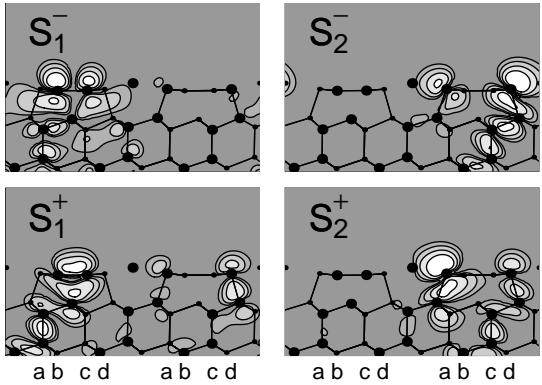}}
\caption{Wave functions (squared) for the four surface states shown in
Fig.~\protect\ref{surfacebands}, evaluated near ${\bf k}$=A. The 
plotting plane is a vertical cut perpendicular to [$\overline{1}$10]
and passing through (from left to right) surface atoms \b, \c, a\, \d\
(projected positions are marked below). Large black circles denote
atoms in the plotting plane, and small circles denote atoms out of
this plane. Contours are logarithmically spaced.
\label{surfacestates}}
\end{figure}

The extent to which this approximate symmetry leads to surface states
with definite parity can be easily evaluated. Given an exact
mirror-plane symmetry and a set of orbitals, $\phi$, associated with
atoms \a\ through \e, the five symmetry-adapted basis functions will
have even ($+$) and odd ($-$) parity:
\begin{equation}
\Phi_1^\pm = \phi_b \pm \phi_c, \; \Phi_2^\pm = \phi_a \pm \phi_d, \;
\Phi_3^+ = \phi_e. \\
\label{symmetryorbitals}
\end{equation}
In Fig.\ \ref{surfacestates} we show the squared modulus of the four
surface states from Fig.\ \ref{surfacebands}, evaluated near the A
point along the [$\overline{1}$10] direction. This figure clearly
shows that \sop\ is a linear combination of $\Phi_1^+$ and $\Phi_3^+$
and thus has overall even parity (the orthogonal state, not shown, is
$\sim$2 eV above the valence edge), while
$\som\approx\Phi_1^-$ has odd parity, as evidenced by the node
at the mirror plane.  Likewise, $\stp\approx\Phi_2^+$ and
$\stm\approx\Phi_2^-$ have overall even and odd parity, respectively.
Having shown that the existence of an approximate surface mirror-plane
symmetry indeed leads to surface states with approximate parity, we
conclude that the matrix element for photoemission from the states
\som\ and \stm\ will be very small.  This explains the absence of any
signal from \stm\ in the ARPES data.  Moreover, we note that the
single peak in the ARPES spectra is quite broad, with a FWHM of
0.7--0.8 eV. We suggest that this broad peak is in fact a
superposition of states \sop\ and \stp. Indeed, excellent fits of the
zone boundary spectrum can be obtained by using two Voigt functions,
representing \sop\ and \stp, separated by 0.4 eV.

The description of the occupied surface states in terms of a
symmetry-adapted basis also leads to a simple explanation for the
stability of the HCC model.  The largest contribution to the \sop\
state is from the $\Phi_1^+$ basis function, which is an even combination
of orbitals separated by one Si bond length---and thus bonding with
respect to \b\ and \c.  The antibonding combination of these orbitals,
$\som\approx\Phi_1^-$, is unoccupied. {\it This suggests the existence
of a true Si double bond, between atoms \b\ and \c, which we propose is
primarily responsible for the stability of the HCC model.}
Indeed, the energy splitting between \sop\ and \som\ is almost 2 eV, far
larger than the bonding-antibonding splitting (less than 0.5 eV) for
the extended Pandey chain.

The Si double bond, although unusual, also occurs in a variety of
molecular systems. In the first known example, the Si=Si bond
was sterically stabilized by two mesityl groups bonded to each Si
\cite{west81}. The simpler disilene molecule, Si$_2$H$_4$, has a
planar structure (see Fig.~\ref{model}) essentially identical to the
arrangement of atoms \a, \b, \c, \d\ in the HCC reconstruction. Here,
the bonds from the substrate to \a\ and \d\ play the role of
steric stabilizers.

We turn now to the appearance of the $M$/Si(111)3\X 1 surface in STM
images. Experimental filled-state images of Li:3\X 1 appear as double
rows of staggered maxima with approximately equilateral spacing;
empty-state images appear as narrow lines with small spurs on both
sides \cite{wan92}. It is at first difficult to reconcile the STM
image of staggered maxima with the model of Fig.~\ref{model}, which
has a mirror plane at the center of each chain.
This, too, has a natural explanation. The 3\X 1 surface
periodicity implies a half-period shift between adjacent chains along
the [$\overline{1}$10] direction. Thus, orbitals $\phi_a$ and $\phi_d$
from two adjacent chains should give rise to a double row of spots
with approximately equilateral spacing. Both
$\stp\approx\phi_a+\phi_d$ and $\stm\approx\phi_a-\phi_d$ lie within
0.5 eV of the valence-band maximum, so that filled-state images with
this bias voltage or larger should appear as described.

\begin{figure}[tbp]
\epsfxsize=5.0cm\centerline{\epsffile{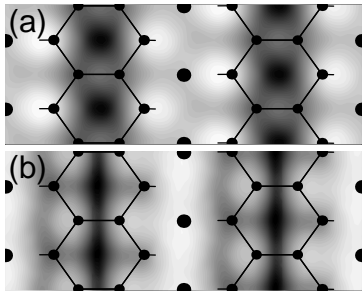}}
\caption{Simulated STM images for the honeycomb chain-channel model of
Li:3\X 1 for (a) filled states and (b) empty states. Black circles
indicate the projected positions of Si and Li atoms in the surface layer.
\label{stm}}
\end{figure}

In Fig.~\ref{stm} we show theoretically simulated constant-current
STM images for the
HCC model, obtained by integrating the local-state density over a
1.5-eV energy window for both filled and empty states. The
filled-state simulated image confirms our qualitative description, and
is in excellent agreement with experimental STM images of Li:3\X 1.
(Note also a small contribution from the \sop\ state, which contributes
to this simulated image because of the large bias voltage chosen.)
The empty-state simulated image is also in excellent agreement with
experiment, appearing as a single row, centered on the Li ions, with
small side spurs. This is easily understood as the superposition of
the unoccupied Li 2$s$ atomic state and the unoccupied \som\ surface
state. We also remark that the registry of the filled- and empty-state
images has been determined experimentally with respect to both the
[$\overline{1}$10] and [11$\overline{2}$] directions \cite{wan92}, and
is in identical to that of the simulated images in Fig.~\ref{stm}.

We return now to the differences between $M$=Li and Ag.  We repeated
the above procedure, starting from the geometry determined by CGM for
the Ag:3\X 1 surface and using an appropriately higher energy cutoff
of 40 Ry for the Ag pseudopotential. In the fully relaxed geometry,
the Ag atom---in contrast to Li---occupies an asymmetric location with
respect to the [$\overline{1}$10] direction and thus breaks this
mirror-plane symmetry.  With respect to the bulk crystal, the
asymmetry is 0.66 \AA, in very good agreement with the diffraction
result of 0.58 \AA \cite{collazo98}.  Moreover, this asymmetry distorts
the triangular spacing observed for Li:3\X 1, and leads instead to a
pairing of atoms \a\ and \d\ from adjacent chains.  Physically, this
asymmetry is driven by the formation of weak Ag-Si bonds, which favors
two-fold coordination of the Ag atom---in contrast to the purely
electrostatic interaction of Li and Si, which favors three-fold
coordination of the Li atom.  Furthermore, this symmetry breaking is
accompanied by a distortion of the surface-layer Si atoms in which the
chain hexagons are rotated by $\sim$7$^\circ$ (these distortions were
not modeled experimentally). This rotation leads to bond strains which
can be partially relieved by alternating the sense of the Ag asymmetry
in adjacent channels. Indeed, we find that this 6\X 1 reconstruction,
which has paired Si atoms but unrotated hexagons, is lower in energy
by 5 meV/\AA$^2$ relative to Ag:3\X 1.  These differences have been
widely observed in comparative STM studies, which indeed find
asymmetrically paired maxima and 6\X 1 periodicity for $M$=Ag, but not
for Li \cite{wan92,carpinelli95}.

Finally, we suggest a reinterpretation of a recent photoemission
experiment for Mg:3\X 1 \cite{an95}. Since Mg is divalent, a band
description of Mg:3\X 1 must be metallic. But high-resolution ARPES
data shows no peak crossing the Fermi level, and on this basis was
interpreted as evidence for a Mott-Hubbard insulator phase
\cite{an95}. Our calculations for Mg:3\X 1 suggest a different
interpretation, motivated by two important results: (1) The fully
relaxed HCC model for Mg:3\X 1 is very similar to Li:3\X 1; in
particular, the mirror plane normal to [11$\overline{2}$] is again
nearly perfect. Thus, by the symmetry argument made above for Li:3\X
1, the matrix element for odd-parity states will vanish when the
incident light is polarized parallel to the mirror plane. (2) The band
structure of Mg:3\X 1 is very similar to Li:3\X 1, with one striking
exception: the \som\ state is pulled down from the conduction manifold
to sit approximately midgap, pinning the Fermi level at $\sim$0.5 eV
above the valence-band edge.  The ARPES data shows two surface states
(labeled $SS_1'$ and $SS_2'$ in Fig. 3 of Ref.\ \cite{an95}) whose
dispersion is in excellent agreement with \stp\ and \sop\ in Fig.\
\ref{surfacebands}. The odd-parity \stm\ state is not observed in the
data; we propose that \som\ is not observed for the same reason. In
other words, Mg:3\X 1 is likely a normal band metal, but appears
insulating as a result of selection rules. Photoemission using
unpolarized light would, we predict, show a peak at the Fermi level.

In summary, we have shown theoretically that the honeycomb
chain-channel reconstruction model for $M$/Si(111)3\X 1 is
energetically far more favorable than any other model proposed to
date. The stability of this model is largely due to the formation of
true Si double bonds in the surface layer. For Li:3\X 1 and Mg:3\X 1,
the existence of a mirror-plane symmetry implies that the
surface states have definite parity---consistent with the findings of
angle-resolved photoemission data. Finally, this model gives a
complete accounting of both filled- and empty-state STM data,
including the important differences between Li:3\X 1 and Ag:3\X 1
surfaces.

Enlightening discussions with R.W. Nunes are acknowledged.
Computational work was supported by a grant of HPC time from the DoD
Major Shared Resource Center \mbox{ASCWP}\@.  S.C.E. is funded by
ONR\@. H.H.W. is funded by NSF under Contract No.\ DMR-9705246. Oak
Ridge National Laboratory is managed by Lockheed Martin Energy
Research Corp.\ for the U.S. Department of Energy under Contract No.\
DE-AC05-96OR22464.

\end{document}